# Interpreting Torsional Oscillator Measurements: Effect of Shear Modulus and Supersolidity

**John D. Reppy, Xiao Mi, Alexander Justin, and Erich J. Mueller**

*Laboratory of Atomic and Solid State Physics, Cornell University, Ithaca, New York, 14853-2501, USA*

*The torsional oscillator is the chief instrument for investigating supersolidity in solid $^4$He. These oscillators can be sensitive to the elastic properties of the solid helium, which show anomalies over the same range of temperature in which the supersolid phenomenon appears. In this report we present a detailed study of the influence of the elastic properties of the solid on the periods of torsional oscillators for the various designs that have been commonly employed in supersolid measurements. We show how to design an oscillator which measures supersolidity, and how to design one which predominantly measures elasticity. We describe the use of multiple frequency TOs for the separation of the elastic and supersolid phenomena.
PACS number: 67.80.bd*

## 1. INTRODUCTION

The first evidence for the supersolid state of solid $^4$He was reported in 2004 by Kim and Chan[1,2]. They observed an anomalous drop in the period of a torsional oscillator containing a solid $^4$He sample when the temperature was lowered below 200 mK. This change in period was interpreted as a supersolid response where a fraction of the moment of inertia of the helium sample decouples from the solid lattice. Kim and Chan found that this period shift was accompanied by a peak in the dissipation. This increased dissipation is not contained in the simplest picture of the supersolid state, where only a frequency independent fraction of the solid $^4$He is expected to decouple from the oscillator.

The experimental picture became more complex with a report, by Day and Beamish[3], of an anomalous decrease of 7% to 15% in the shear modulus of the solid as the temperature was increased from 20 to 500 mK. The largest portion of this reduction occurs over the same temperature region as the supersolid behavior reported by Kim and Chan. The Day-Beamish measurements confirmed an earlier observation by Paalanen, Bishop, and Dail[4]. Paalanen et al. found that $\mu_{low}/\mu_{0.5K} = 1.40$, where $\mu_{low}$ and $\mu_{0.5K}$ are the shear moduli at low temperature and at 0.5 K. In more recent

**John D. Reppy, Xiao Mi, Alexander Justin, and Erich J. Mueller**

measurements, Mukharsky, Penzev, and Varoquaux[5] have seen a ratio as large as $\mu_{low}/\mu_{0.5K} = 1.58$.

There is an almost complete correspondence between the various aspects of supersolid behavior and the phenomena associated with the elastic anomaly, including sensitivity to $^3$He impurity level, dissipative effects, sensitivity to strain or velocity amplitude, and hysteresis.

An important question is to what extent the observed frequencies and quality factors are due to supersolidity or elasticity. This question has been addressed by a number of workers in the field: initially with a finite element approach by the Alberta and Penn State group of West, Syshchenko, Beamish, and Chan[6] and by the analytic solution of simple model systems[7]. Recently, Maris and Balibar[8] have returned to this problem, employing a perturbation approach. The general conclusion has been that for most torsional oscillator designs, period shifts due to the elastic anomaly are relatively small. However, it is possible to design an oscillator where the effects of the elastic anomaly are considerably enhanced.

It is our aim in this paper to consider the elastic effects for a number of different torsional oscillator designs that have been employed in our supersolid research program at Cornell. Our approach will be analytic using simplified models for the mechanical oscillators. We believe that this approach will provide physical insight that is lost in the application of the finite element approach. We will proceed by considering a sequence of increasingly complex torsional oscillator designs leading to an analysis of our double and triple oscillator structures.

Section 2 explores elastic effects in a simple torsional oscillator. Section 3 extends these results to the case of an oscillator filled with solid helium. Section 4 analyzes more complicated compound oscillators. Section 5 presents conclusions.

## 2. A SIMPLE TORSIONAL OSCILLATOR

A typical torsional oscillator is illustrated in Fig. 1.

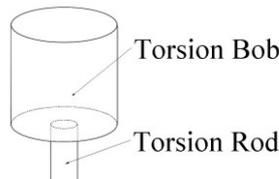

Fig. 1. A Simple Torsional Oscillator

It consists of a solid torsion bob or head with radius $r$, length $h$, and density $\rho$. The moment of inertia about the cylindrical axis is $I = \pi \rho h r^4 / 2$. The bob

# Interpreting Torsional Oscillator Measurements

is mounted on torsion rod with a static torsion constant $k = \pi\mu d^4 / 2l$, where $l$ is the length, $\mu$ the shear modulus, and $d$ the radius. The lower end of the torsion rod is rigidly fixed to a very large mass. The torque provided to the torsion head by the torsion rod is $\tau = -k\theta$, where $\theta(t) = \theta_0 \sin(\omega t)$ is the angular displacement of the torsion bob and the natural period for oscillation is $P = 2\pi\sqrt{I/k}$.

In this treatment we have made two approximations: first, that the torsion bob is "completely rigid," i.e., it has infinite shear modulus; second, that the density of the torsion rod is zero. The assumptions of a completely rigid cell and a torsion rod with zero mass have been standard in the analysis of supersolid torsional oscillator experiments since the early experiments of Kim and Chan.

In this section we consider the consequences of relaxing these assumptions. In the interests of simplicity we shall assume that the oscillator head and torsion rod are made of the same material.

## 2a. Torsion Rod with $\rho \neq 0$

We model the torsion rod as a uniform cylinder of length $l$, radius $d$, and shear modulus $\mu$, and with its base fixed to a large rigid mass and its top attached to the bob. The differential equation governing oscillations about the axis of the cylinder is

$$d^2\theta/dz^2 - (\mu/\rho)d^2\theta/dt^2 = 0, \tag{1}$$

where the coordinate $z$ is along the axis of the cylinder with the origin at the base and $\theta(z,t) = \theta(z)\sin(\omega t)$ is the angular displacement of the cylinder. The boundary conditions are $\theta(z=0) = 0$ and $(d\theta/dz)_{z=l} = -\tau_{bob}/(\pi\mu d^4/2)$ where $\tau_{bob}\sin(\omega t)$ is the torque on the rod from the bob. The resulting angular profile is $\theta(z) = \theta_0 \sin(2\pi z/\lambda)/\sin(2\pi l/\lambda)$ where $\theta_0$ is the maximum amplitude of angular oscillation at the bob, and the wavelength is set by $\lambda = (2\pi/\omega)(\mu/\rho)^{1/2} = P(\mu/\rho)^{1/2}$ where $P$ is the period of oscillation.

Differentiating the angular profile and comparing with the boundary conditions give $\tau_{bob} = -(\pi\mu d^4/2)(d\theta/dz) = -k_0(2\pi l/\lambda)\cot(2\pi l/\lambda)\theta_0$ where $k_0 = (\pi\mu d^4/2l)$ is the usual static expression for the torsion constant of the rod. One can define an effective frequency dependent torsion constant. We see that the finite mass of the rod can be absorbed into a frequency dependent torsional constant:

$$k(\rho,\omega) = -\tau_{bob}/\theta_0 = k_0(2\pi l/\lambda)\cot(2\pi l/\lambda). \tag{2}$$

The rod's mass density appears in $\lambda$.

**John D. Reppy, Xiao Mi, Alexander Justin, and Erich J. Mueller**

High-$Q$ torsion rods are commonly composed of heat treated BeCu with a shear modulus $\mu = 5.3 \times 10^{11}$ dyne/cm$^2$ and a density $\rho = 8.23$ gm/cm$^3$. Taking our rod to have a length 2.0 cm, oscillating at 1 kHz, the term $2\pi l / \lambda = 4.9 \times 10^{-2}$. Since this term is small, we shall expand the cotangent as a Taylor series in $(2\pi l / \lambda)$:

$$k(\rho,\omega) \approx k_0 \left[1 - 1/3(2\pi l / \lambda)^2\right] = k_0 \left[1 - 1/3\left(l^2 \omega^2 \rho / \mu\right)\right]. \quad (3)$$

Thus, there is only a small reduction in the effective torsion constant resulting from non-zero density. In the specific case of the BeCu torsion rod with length $l = 2$ cm, the fractional change is $\delta k / k = 1/3\left(l^2 \omega^2 \rho / \mu\right) \approx 1.43 \times 10^{-4}$. This change can generally be neglected.

## 2b. Torsion Head with $\mu \neq \infty$

The torsion head can be treated in much the same way as the torsion rod. We model it as a uniform cylinder of length $h$ (so the top is at $z_{end} = L = l + h$) and radius $r$. There is no torque at the top of the oscillator head and thus the boundary condition is $(d\theta / dz)_{z=z_{end}} = 0$. We therefore conclude $\theta(z) = \theta_0 \cos(2\pi(L-z)/\lambda) / \cos(2\pi h/\lambda)$. The bob exerts a torque of $\tau_{bob} = \left(\pi \mu r^4 / 2\right)(d\theta / dz) = -I_0 \omega^2 (\lambda / 2\pi h) \tan(2\pi h / \lambda) \theta_0$ on the torsion rod, where $I_0 = \pi r^4 \rho h / 2$ is the moment of inertial of a rigid cylinder. The finite shear modulus of the cylinder leads to a frequency dependent moment of inertia

$$I(\mu,\omega) = -\tau_{bob} / (\omega^2 \theta_0) = I_0 (\lambda / 2\pi h) \tan(2\pi h / \lambda). \quad (4)$$

Once again we can expand the tangent, and to leading order

$$I_{eff} \approx I_0 \left[1 + \left(h^2 \omega^2 \rho / \mu\right)/3\right]. \quad (5)$$

The effective moment of inertia will increase with decreasing shear modulus, thus leading to a decrease in frequency. In the case of a BeCu torsion head with height $h = 2$ cm oscillating at 1 kHz, the fractional increase in the effective moment of inertia going from the rigid to non-rigid torsion bob is again $\left(h^2 \omega^2 \rho / \mu\right)/3 = 1.43 \times 10^{-4}$. Thus, the error introduced by assuming that the torsion bob is completely rigid is small and can be safely neglected in most applications. The physically important aspect arising from this treatment is the fact that the effective moment of inertia increases as the square of oscillation frequency and also increases as the shear modulus is reduced. This behavior will also be a feature of our calculations for torsional oscillators containing solid $^4$He samples.

## Interpreting Torsional Oscillator Measurements

## 3. TORSIONAL OSCILLATORS CONTAINING SOLID $^4$HE SAMPLES

In this section we consider the effects of the changes in the mass and elastic properties of solid $^4$He on the resonance frequencies of single-mode helium filled torsional oscillators with different internal geometries (see Figs. 2 and 3). We treat $^4$He as a linear, isotropic medium. Following our discussion of the simple oscillator, the main effect of the finite shear modulus of the solid helium is to modify the moment of inertia of the oscillator in a frequency dependent way. As before, the effective moment of inertia will increase with decreasing shear modulus. We also treat the oscillator walls as rigid. The boundary condition on the surfaces of the solid $^4$He will therefore correspond to rigid body rotation of the container with a fixed amplitude $\theta_0$. In section 4, we will consider more complicated geometries. Generically the interior of the helium will move with a larger amplitude than the boundaries.

The motion of the oscillator is clearest in complex notation, and we write $\theta = \theta_0 e^{i\omega t}$. The physical angle of the bob is the imaginary part of $\theta$. The forces on the bob come from the torsion rod and the helium, allowing us to write $-I\omega^2\theta + k\theta = \tau$, where $\tau = -\tau_{He} = \omega^2 I_{eff} \theta$ is the torque exerted on the bob by the helium. If the solid helium was completely rigid, then $I_{eff}$ would simply be the moment of inertia of the helium.

We introduce a "susceptibility", $\chi(\omega) = \tau(\omega)/(\theta_0 e^{i\omega t}) = -\omega^2 I_{eff}$ which is the amplitude of the reaction torque divided by the amplitude of the oscillator's angular displacement. The equation of motion is then:

$$I\omega^2 - \chi(\omega) - k = 0. \qquad (6)$$

For a typical supersolid torsional oscillator, the ratio of the moment of inertia of the helium sample to the moment of inertia of the container is $I_{He}/I \approx 10^{-3}$. Therefore, $\omega^2 I_{eff} \ll \omega^2 I$ and $\chi(\omega) = -\omega^2 I_{eff}$ can be treated as a small perturbation for finding the resonance frequency. We shall now calculate $I_{eff}$ for several sample geometries.

### 3a. Cylindrical Geometry

In this design, a hollow cylindrical cavity in the interior of the torsion bob is filled with solid $^4$He. We adopt a cylindrical coordinate system aligning the axis of the cylinder with the *z*-axis. The length of the cylindrical cavity is $z_0$ and its radius $r_0$. Choosing the origin at the center of the cylinder, the bottom and top are located at $z = -z_0/2$ and $z = z_0/2$. At the frequencies and dimensions of typical supersolid torsional oscillators, we may assume

**John D. Reppy, Xiao Mi, Alexander Justin, and Erich J. Mueller**

that the solid is incompressible, i.e. $\nabla \cdot \boldsymbol{u} = 0$, and the Navier-Cauchy equation reduces to the following form:

$$\nabla^2 (\nabla \times \boldsymbol{u}) - \frac{\rho}{\mu} \frac{\partial^2 (\nabla \times \boldsymbol{u})}{\partial t^2} = 0, \tag{7}$$

where $\mu$ is the shear modulus of solid $^4$He and $\rho$ is its density. Assuming the solution is of the form $\boldsymbol{u} = Z(z)u(r)e^{i\omega t}\boldsymbol{e}_\phi$, equation (13) decouples into two parts:

$$\frac{d^2 Z}{dz^2} = -k_0^2 Z \text{ and} \tag{8}$$

$$r^2 \frac{d^2 R}{dr^2} + r \frac{dR}{dr} + \left[ (\frac{\omega^2}{\beta^2} - k_0^2)r^2 - 1 \right] R = 0, \tag{9}$$

where $k_0$ is a constant and could take on any value. We will first solve for $\boldsymbol{u}$ and hence $\chi(\omega)$ in the two limiting cases, those of a "pancake" or disc geometry where $z_0 \ll r_0$ with the effects of the sidewalls neglected; and a "thin-rod" or infinite cylinder geometry where $r_0 \ll z_0$ with the effects of the top/bottom walls neglected. We will then solve for the general case where $z_0$ and $r_0$ are comparable in size. Fig. 2 illustrates the three cases we will be considering.

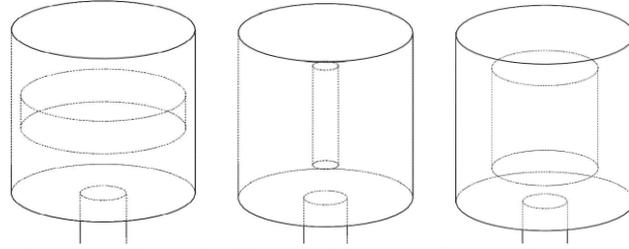

a. "pancake"     b. "thin-rod"     c. finite cylinder
Fig. 2. Oscillators with different inner cylindrical cavities

In the limit $z_0 \ll r_0$, we need only to solve equation (7), subject to the boundary condition $\boldsymbol{u}(z = \pm z_0/2) = r\theta_0 e^{i\omega t}\boldsymbol{e}_\phi$. This could be satisfied by setting $k_0 = \omega / \beta$. $R$ is then a linear function in $r$ whereas $Z$ is sinusoidal. The overall solution is:

$$\boldsymbol{u} = \frac{r\theta_0 \cos\left(z\omega\sqrt{\rho/\mu}\right)}{\cos\left((z_0/2)\omega\sqrt{\rho/\mu}\right)} e^{i\omega t}\boldsymbol{e}_\phi. \tag{10}$$

The displacement then determines the stress tensor component, $\sigma_{z\phi}$, and thereby $\chi(\omega)$ through integrating $r\sigma_{z\phi}$ over the top and bottom surfaces:

## Interpreting Torsional Oscillator Measurements

$$\chi(\omega) = \frac{4\pi}{\theta_0 e^{i\omega t}} \int_0^{r_0} \sigma_{z\phi}\big|_{z=\pm 1/2 z_0} r^2 dr = \frac{4\pi}{\theta_0 e^{i\omega t}} \int_0^{r_0} \mu \frac{\partial u_\phi}{\partial z}\bigg|_{z=\pm 1/2 z_0} r^2 dr. \quad (11)$$

The susceptibility term for the pancake oscillator is:

$$\chi(\omega) = -\pi r_0^4 \omega \sqrt{\rho\mu} \tan\left(\frac{1}{2} z_0 \omega \sqrt{\rho/\mu}\right) \approx -I_{eff} \omega^2 \quad (12)$$

where $I_{eff} = I_{He}\left(1 + \frac{1}{12} z_0^2 \omega^2 \frac{\rho}{\mu}\right)$ and $I_{He}$ is the moment of inertia of the helium sample. The Taylor expansion is justified by the fact that $z_0 \ll r_0$. For a value of $z_0 = 1$ mm, the term $\frac{1}{12} z_0^2 \omega^2 \frac{\rho}{\mu} \approx 4.4 \times 10^{-5}$.

The next limit we will consider is the "infinite" cylinder case with $r_0 \ll z_0$, where we need only to be concerned with the boundary condition $\mathbf{u}(r = r_0) = r_0 \theta_0 e^{i\omega t} \mathbf{e}_\phi$. This limiting case has also been discussed by I. Iwasa[9] reaching an identical conclusion. The displacement field is:

$$\mathbf{u} = \frac{r_0 \theta_0}{J_1(r_0 \omega \sqrt{\rho/\mu})} J_1(r\omega\sqrt{\rho/\mu}) e^{i\omega t} \mathbf{e}_\phi. \quad (13)$$

$J_n$ is Bessel function of order $n$. The relevant stress tensor component is determined to be:

$$\sigma_{r\phi} = \mu\left(\frac{du_\phi}{dr} - \frac{u_\phi}{r}\right) = -\frac{r_0 \theta_0 \omega \sqrt{\rho\mu}}{J_1(r_0\omega\sqrt{\rho/\mu})} J_2(r\omega\sqrt{\rho/\mu}) e^{i\omega t}. \quad (14)$$

The resulting susceptibility is then determined by integrating $r\sigma_{z\phi}$ over the cylindrical boundary surface at $r = r_0$:

$$\chi(\omega) = -2\pi\mu z_0 r_0^3 \omega \sqrt{\frac{\rho}{\mu}} \frac{J_2(r_0\omega\sqrt{\rho/\mu})}{J_1(r_0\omega\sqrt{\rho/\mu})} \approx -I_{eff}\omega^2. \quad (15)$$

where $I_{eff} = I_{He}\left(1 + \frac{1}{24} r_0^2 \omega^2 \frac{\rho}{\mu}\right)$. The form of the effective moment of inertia is very similar to the "pancake" oscillator. For a value of $r_0 = 1$ cm, the term $\frac{1}{24} r_0^2 \omega^2 \frac{\rho}{\mu} \approx 2.2 \times 10^{-3}$.

An estimate for the apparent Non-Classical Rotational Inertia Fraction (NCRIF) is given by the fractional change in the effective moment of inertia of the sample between low and high temperature by:

**John D. Reppy, Xiao Mi, Alexander Justin, and Erich J. Mueller**

$$\frac{\Delta I_{\text{eff}}}{I_{He}} = -\frac{r_0^2 \omega^2 \rho}{24 \mu_{\text{low}}}\left(\frac{\Delta \mu}{\mu_{\text{high}}}\right) \approx -2.2 \times 10^{-3}\left(\frac{\Delta \mu}{\mu_{\text{high}}}\right) = 2.2 \times 10^{-3} \delta \quad (16)$$

where $\mu_{\text{low}}$ and $\mu_{\text{high}}$ are the low and high temperature values of the shear modulus, $\Delta\mu = \mu_{\text{high}} - \mu_{\text{low}}$ and $\delta = [\mu_{\text{low}} / \mu_{\text{high}} - 1]$.

In the general case where $z_0$ is comparable to $r_0$, the boundary conditions at the cylindrical wall and those at the top and bottom surfaces must be considered. Therefore we need to solve (7) subject to $\boldsymbol{u}(z = \pm z_0/2) = r\theta_0 e^{i\omega t} \boldsymbol{e}_\phi$ and $\boldsymbol{u}(r = r_0) = r_0\theta_0 e^{i\omega t} \boldsymbol{e}_\phi$. The addition of the boundary condition at the top and bottom surfaces reduces $I_{\text{eff}}$ below the infinite cylinder value. The general approach to solving this type of problem involves superimposing the solutions to two separate problems in which the first solution satisfies $\boldsymbol{u}(z = \pm z_0/2) = 0$ and $\boldsymbol{u}(r = r_0) = r_0\theta_0 e^{i\omega t} \boldsymbol{e}_\phi$, and the second solution satisfies $\boldsymbol{u}(z = \pm z_0/2) = r\theta_0 e^{i\omega t} \boldsymbol{e}_\phi$ and $\boldsymbol{u}(r = r_0) = 0$. The separation constant $k_0$ takes on an infinite number of quantized values and the solution emerges as a sum of two infinite series:

$$\boldsymbol{u} = \sum_{n \text{ odd}} A_n I_1\left(r\sqrt{(n\pi/z_0)^2 - \omega^2 \rho/\mu}\right)\cos(n\pi z/z_0)e^{i\omega t}\boldsymbol{e}_\phi$$
$$+ \sum_{\text{all } n} B_n J_1(\alpha_{1n} r/r_0)\cosh\left(z\sqrt{(\alpha_{1n}/r_0)^2 - \omega^2 \rho/\mu}\right)e^{i\omega t}\boldsymbol{e}_\phi \quad (17)$$

$I_n$ is a modified Bessel function of order $n$, and $\alpha_{1n}$ is the $n^{\text{th}}$ root to $J_1$. Using Fourier's method, we find that the coefficients $A_n$ and $B_n$ are:

$$A_n = \frac{4r_0\theta_0}{n\pi}\frac{1}{I_1\left(r_0\sqrt{(\pi n/z_0)^2 - \omega^2\rho/\mu}\right)}. \quad (18)$$

$$B_n = \frac{2r_0\theta_0}{\alpha_{1n}J_2(\alpha_{1n})}\frac{1}{\cosh\left(\frac{z_0}{2}\sqrt{(\alpha_{1n}/r_0)^2 - \omega^2\rho/\mu}\right)}. \quad (19)$$

The susceptibility is then computed through integrating $rF$, where $F$ is the local force density, over the volume of the cavity:

$$\chi(\omega) = \frac{1}{\theta_0 e^{i\omega t}}\int_{\text{Volume}} rF\, dV = \frac{\rho}{\theta e^{i\omega t}}\int_{\text{Volume}} r\frac{d^2 u_\phi}{dt^2}\, dV = -I_{\text{eff}}\omega^2. \quad (20)$$

**Interpreting Torsional Oscillator Measurements**

$$I_{eff} = I_{He} \left[ \begin{array}{l} \dfrac{8}{\pi r_0^2} \sum_{n \, odd} A_n \dfrac{1}{n\sqrt{(\pi n/z_0)^2 - \omega^2 \rho/\mu}} I_2\left(r_0 \sqrt{(\pi n/z_0)^2 - \omega^2 \dfrac{\rho}{\mu}}\right) + \\ \dfrac{8}{r_0 z_0} \sum_{all \, n} B_n \dfrac{J_2(\alpha_{1n})}{\alpha_{1n} \sqrt{(\alpha_{1n}/r_0)^2 - \omega^2 \rho/\mu}} \sinh\left(\dfrac{1}{2} z_0 \sqrt{(\alpha_{1n}/r_0)^2 - \omega^2 \dfrac{\rho}{\mu}}\right) \end{array} \right]$$

This series converges with increasing $n$ and can be evaluated conveniently with the aid of a computer.

### 3b. Annular Geometry

Annular cells have been used in a variety of experiments[2,9,10,11,12]. The original motivation for adopting a narrow annular geometry arose from critical velocity studies[2] which took advantage of the more limited range of velocities. In many designs there is an elastic connection such as a welding or epoxy joint (Fig. 5a) between the inner and outer moments of inertia and a small relative motion is possible. We shall consider first, however, a perfectly rigid annulus as shown in Fig. 3 where no relative rotational motion is possible between the inner and outer walls of the sample volume.

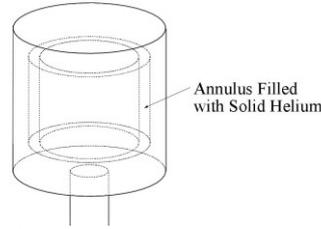

Fig. 3. Rigid Annular Torsional Oscillator

In the case of a narrow annular geometry, $\Delta r_0 \ll r_0$ and we could treat the problem in 2-D Cartesian coordinates by "unrolling" the annulus. Setting the average radius of the annulus to be $r_0$, we situate the two side walls of the annulus at $y = \pm \Delta r_0/2$, where $\Delta r_0$ is the width of the annulus. The displacement is $\boldsymbol{u} = u(y)e^{i\omega t}\boldsymbol{e}_x$, where:

$$\partial^2 u / \partial y^2 = (\rho/\mu) \partial^2 u / \partial t^2 . \qquad (21)$$

Subject to the boundary condition that $u = r_0 \theta$ at $y = \pm \Delta r_0/2$ the solution is:

$$\boldsymbol{u} = \dfrac{r_0 \theta_0 \cos\left(y\omega\sqrt{\rho/\mu}\right)}{\cos\left(\dfrac{1}{2}\Delta r_0 \omega \sqrt{\rho/\mu}\right)} e^{i\omega t} \boldsymbol{e}_x . \qquad (22)$$

The torque exerted by solid $^4$He on the oscillator is:

John D. Reppy, Xiao Mi, Alexander Justin, and Erich J. Mueller

$$\tau = \int r_0 \sigma_{yx}\big|_{y=\pm\frac{1}{2}\Delta r_0} ds = -4\pi r_0^3 z_0 \omega\sqrt{\rho\mu}\tan\left(\frac{1}{2}\Delta r_0 \omega\sqrt{\rho/\mu}\right)\theta_0 e^{i\omega t}, \quad (23)$$

where $z_0$ is the height of the annulus. After expanding the tangent function and keeping the lowest significant term, we have:

$$\tau/\theta_0 e^{i\omega t} = \chi(\omega) \approx -I_{eff}\omega^2, \text{ where } I_{eff} = I_{He}\left(1+\frac{1}{12}(\Delta r_0)^2 \omega^2 \frac{\rho}{\mu}\right). \quad (24)$$

The form of the susceptibility is again very similar to (12) and (15). For a value of $\Delta r_0$ = 0.5 mm, the term $\frac{1}{12}(\Delta r_0)^2 \omega^2 \frac{\rho}{\mu} \approx 1.1\times10^{-5}$. In all the oscillators discussed in this section, $\mu$ contributes very little to the effective moment of inertia of the helium sample, and therefore changes in $\mu$ due to the elastic anomaly will lead to only small changes in the resonance frequency. This will not be necessarily true in all cases, as we shall explore in Section 4.

### 3c. Frequency Shifts

We now determine the frequency shifts caused by the solid $^4$He shear modulus anomaly in the "pancake," "thin-rod" and annular oscillators. The case for the cylindrical oscillator where the length and radius are comparable cannot be put in closed forms, but can be easily evaluated by computer. In all cases the susceptibility terms are conveniently expressed as:

$$\chi(\omega) \cong -I_{He}\omega^2\left[1+\frac{RD^2\omega^2\rho}{\mu}\right] = -\omega^2 I_{eff}, \quad (25)$$

where $R$ = 1/12 for the pancake and annular geometries and 1/24 for the infinite cylinder case. $D$ is the height $z_0$ in the pancake geometry, the radius $r_0$ in the infinite cylinder geometry, and $D = \Delta r_0$ for the annular geometry. The first order solution to (6) is:

$$\omega^2 \approx k/I + \chi\left(\sqrt{k/I}\right)/I. \quad (26)$$

In the event of shear-stiffening, the variation in the resonance period P is:

$$\frac{\delta P}{\delta \mu} = -\frac{P^3}{8\pi^2}\frac{\delta(\omega^2)}{\delta\chi\left(\sqrt{k/I}\right)}\frac{\delta\chi\left(\sqrt{k/I}\right)}{\delta\mu} = -\frac{P^3 k^2 RD^2 I_{He}\rho}{8\pi^2 I^3 \mu^2}. \quad (27)$$

In the event of Non-Classical Rotational Inertia (NCRI), i.e. supersolid, the variation in the resonance period is:

$$\frac{\delta P}{\delta I_{He}} = -\frac{P^3}{8\pi^2}\frac{\delta(\omega^2)}{\delta\chi\left(\sqrt{k/I}\right)}\frac{\delta\chi\left(\sqrt{k/I}\right)}{\delta I_{He}} = \frac{kP^3}{8\pi^2 I^2}. \quad (28)$$

## Interpreting Torsional Oscillator Measurements

Lastly, although the effect of shear stiffening on the periods of these oscillators is small, it can still lead to an apparent NCRI fraction $\Delta I_{eff} / I_{He}$, since $I_{eff}$ changes with a changing $\mu$. John Beamish has kindly provided us with a representative set of the Day-Beamish shear modulus data. For this data set $\mu = 1.642 \times 10^8$ dyne/cm$^2$ at the lowest temperature and decreases by about 7% to $\mu = 1.528 \times 10^8$ dyne/cm$^2$ at 0.5 K.

Using these data, we have plotted, at a frequency of 1000 Hz, the apparent NCRI fraction arising from changes in $\mu$ alone for the four geometries discussed. Although the change in the shear modulus is only 7% for the data set we are using, much larger changes, up to 40%[5] and even approaching 60%[6] have been reported and would lead to period shifts almost an order of magnitude larger than those shown in Fig. 4.

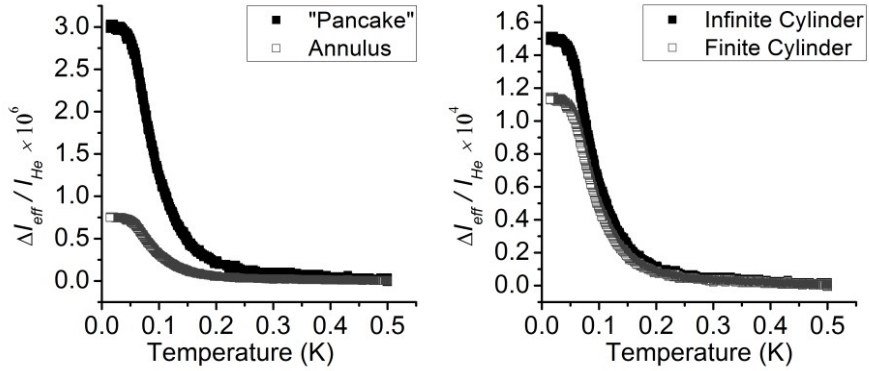

Fig. 4. Apparent NCRI fraction due to shear stiffening for different geometries ($z_0 = 1$ mm for the "Pancake" geometry, $\Delta r_0 = 0.5$ mm for annular geometry, $r_0 = 1$ cm for infinite cylinder geometry, $r_0 = 1$ cm and $z_0 = 2$ cm for the finite cylindrical geometry.)

## 4. DOUBLE OSCILLATORS

Distinguishing between shear stiffening and NCRI is difficult with a single-mode oscillator, since both raise the resonance frequency. However, with a dual-mode oscillator, the two phenomena can be separated, since the ratios of frequency shifts of the two modes are different for the two phenomena. We shall first consider a double oscillator constructed by mounting a single oscillator, with one of the geometries discussed in the previous section, on a dummy oscillator. In a more comprehensive treatment, we shall consider a triple oscillator design where a third degree of freedom is provided by an additional torsion bob coupled to the motion of an oscillator that is mounted on another dummy oscillator, through the elasticity of the solid $^4$He sample.

**John D. Reppy, Xiao Mi, Alexander Justin, and Erich J. Mueller**

**4a. Simple Double Oscillator**

We first construct a dummy oscillator with torsion constant $k_3$ and moment of inertia $I_3$. The subscript notation has been chosen in anticipation of the triple oscillator case. One end of the torsion rod for the dummy oscillator is attached to the bottom of the massive mixing chamber and is assumed to be stationary. The oscillator containing the $^4$He sample, with torsion constant $k_2$ and moment of inertia $I_2$, is then mounted on the dummy oscillator. The equations of motion then read:

$$\begin{pmatrix} k_3 + k_2 - \omega^2 I_3 & -k_2 \\ -k_2 & k_2 - \omega^2 I_2 + \chi(\omega) \end{pmatrix} \begin{pmatrix} \theta_3 \\ \theta_2 \end{pmatrix} = 0. \quad (29)$$

where $\theta_3$ and $\theta_2$ are the angular displacements of the dummy oscillator and helium filled oscillator. In our triple oscillator discussion, $\theta_1$ will represent the angular displacement of a third element, $I_1$. The term $\chi(\omega)$ has the same form and physical meaning as developed in the previous section. The normal mode frequencies are obtained by setting the determinant of the above 2 × 2 matrix to zero. In the absence of a $^4$He sample, $\chi(\omega) = 0$, and the high (+) and low (-) mode frequencies denoted by $\omega_\pm$ are given by:

$$\omega_\pm^2 = \frac{I_2 k_2 + I_2 k_3 + I_3 k_2}{2 I_2 I_3} \left( 1 \pm \sqrt{1 - \frac{4 I_2 I_3 k_2 k_3}{(I_2 k_2 + I_2 k_3 + I_3 k_2)^2}} \right). \quad (30)$$

Treating $\chi(\omega)$ as a small perturbation, the solutions for the squares of the angular frequencies to linear order in $\chi(\omega)$ are:

$$\omega_{High}^2 = \omega_+^2 + \chi(\omega_+) \frac{\omega_+^2 I_3 - k_3 - k_2}{I_3 I_2 (\omega_+^2 - \omega_-^2)} \text{ and } \omega_{Low}^2 = \omega_-^2 + \chi(\omega_-) \frac{k_3 + k_2 - \omega_-^2 I_3}{I_3 I_2 (\omega_+^2 - \omega_-^2)}. \quad (31)$$

Knowing the dimensions of the oscillators, one could use the forms of $\chi(\omega)$ for specific geometries to predict the magnitudes of frequency shifts for the two different modes. It is particularly useful to look at the ratio of period shifts, $\delta P_{High} / \delta P_{Low}$, at the two modes in the event of NCRI and shear stiffening. The ratios can be obtained through application of the chain rule $\delta P_{High/Low} / \delta X = (\delta P_{High/Low} / \delta \chi(\omega_{+/-}))(\delta \chi(\omega_{+/-}) / \delta X)$, where $X$ is either $I_{He}$ or $\mu$. The ratios are:

$$\left. \frac{\delta P_{High}}{\delta P_{Low}} \right|_{NCRI} = \frac{\delta P_{High} / \delta I_{He}}{\delta P_{Low} / \delta I_{He}} \approx \frac{P_+}{P_-} \frac{\omega_+^2 I_3 - k_3 - k_2}{k_3 + k_2 - \omega_-^2 I_3}, \quad (32)$$

$$\left. \frac{\delta P_{High}}{\delta P_{Low}} \right|_{Shear} = \frac{\delta P_{High} / \delta \mu}{\delta P_{Low} / \delta \mu} \approx \frac{P_-}{P_+} \frac{\omega_+^2 I_3 - k_3 - k_2}{k_3 + k_2 - \omega_-^2 I_3}. \quad (33)$$

Therefore, the period shift ratio for each scenario is given by:

# Interpreting Torsional Oscillator Measurements

$$\frac{\left[\delta P_{High}/\delta P_{Low}\right]_{NCRI}}{\left[\delta P_{High}/\delta P_{Low}\right]_{Shear}} = \frac{P_{High}^2}{P_{Low}^2} \qquad (34)$$

This equation is applicable to all geometries discussed in Section 3, providing a convenient means for distinguishing between the effects of NCRI and shear-stiffening.

### 4b. Compound Oscillators

The compound oscillators shown below in Fig. 5 will have the same equations of motion. The notation we have adopted is again meant to be consistent with the notation to be used later for the triple oscillator case. Here the torsion constant $k_0$ is the torsion constant due to the epoxy for the first oscillator (Fig. 5a) and to the torsion constant of the internal torsion rod supporting the internal moment of inertia $I_1$ in the case of the "Floating-core" oscillator (Fig. 5b).

The structure of the double oscillator considered here resembles that of a single oscillator with annular geometry. However, the internal torsion bob is free to rotate with respect to the outer oscillator and therefore constitutes an additional oscillator. This freedom can be provided by a well-defined torsion rod as in Fig. 5a, or epoxy/welded joints as in Fig. 5b.

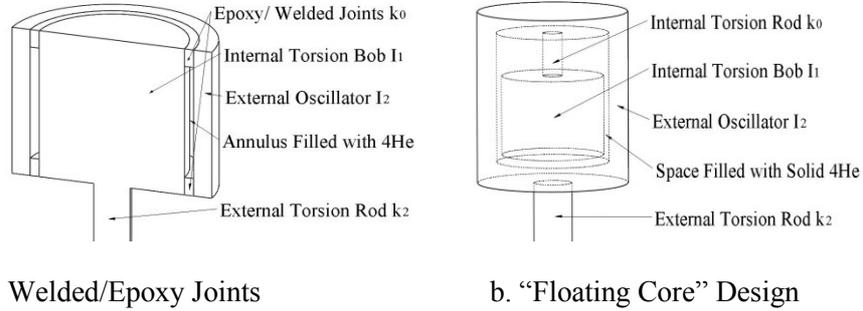

   a. Welded/Epoxy Joints            b. "Floating Core" Design
Fig. 5. Annular Compound Oscillators

For a compound oscillator of the above designs, the equations of motion read:

$$\begin{pmatrix} k_0 + k_2 - \omega^2 I_2 - \chi_{11}(\omega) & -\chi_{12}(\omega) - k_0 \\ -\chi_{21}(\omega) - k_0 & k_0 - \omega^2 I_1 - \chi_{22}(\omega) \end{pmatrix} \begin{pmatrix} \theta_2 \\ \theta_1 \end{pmatrix} = 0. \qquad (35)$$

Here $\theta_2$ and $\theta_1$ are the angular displacements of the external oscillator and the internal torsion bob respectively. Typically, the gap between the internal torsion bob and outer oscillator is small. The primary contribution to the susceptibility terms arises from the solid $^4$He in the annular region. Also, the narrowness of the annulus enables us to solve for the displacement $\boldsymbol{u}$ of solid

**John D. Reppy, Xiao Mi, Alexander Justin, and Erich J. Mueller**

$^4$He in the same Cartesian coordinates as in Section 3b. In this case, however, the boundary conditions are $\mathbf{u} = r_0\theta_2 e^{i\omega t}\mathbf{e}_x$ at $y = +\Delta r_0/2$ and $\mathbf{u} = r_0\theta_1 e^{i\omega t}\mathbf{e}_x$ at $y = -\Delta r_0/2$, where $\theta_2$ is not necessarily equal to $\theta_1$. The solution is:

$$\mathbf{u} = \left[ r_0\theta_2 \frac{e^{ik(1/2\Delta r - y)} - e^{-ik(1/2\Delta r - y)}}{e^{ik\Delta r} - e^{ik\Delta r}} + r_0\theta_1 \frac{e^{ik(1/2\Delta r + y)} - e^{-ik(1/2\Delta r + y)}}{e^{ik\Delta r} - e^{ik\Delta r}} \right] \mathbf{e}_x, \qquad (36)$$

where $k = \omega\sqrt{\rho/\mu}$. Solving for the susceptibilities and Taylor-expanding:

$$\chi_{11} = \chi_{22} = -\left( k_\mu + \frac{I_{He}}{3}\omega^2 \right) + \frac{I_{He}}{2}\omega^2 = -\left( k_\mu - \frac{I_{He}}{6}\omega^2 \right), \qquad (37)$$

$$\chi_{12} = \chi_{21} = \left( k_\mu + \frac{I_{He}}{3}\omega^2 \right), \qquad (38)$$

where $k_\mu = 2\pi r_0^3 z_0 \mu / \Delta r_0$ is the torsion constant contributed by solid $^4$He in the zero frequency limit. Supposing $I_{He} = 0.2$ gm-cm$^2$, $\omega = 2\pi \times 1000$ Hz, $\mu = 1.5 \times 10^8$ dyne/cm$^2$, $r_0 = 1$cm. $\Delta r_0 = 0.5$ mm and $z_0 = 2$ cm, we have $k_\mu = 3.8 \times 10^{10}$ dyne-cm and $I_{He}\omega^2 = 2.0 \times 10^6$ dyne-cm, which differ by four orders of magnitude. Hence, for a narrow annulus, the term $k_\mu$ dominates in all four susceptibility terms. To first order, the helium sample contributes both as a torsion rod that couples $I_1$ with $I_2$ and as an additional mass loading of $(1/2)I_{He}$ to both $I_1$ and $I_2$.

We shall consider first the oscillator configuration with epoxy joints (Fig. 5a). We estimate the torsion constant $k_0$ using a shear modulus value for epoxy, $\mu_{epoxy} = 3.0 \times 10^{10}$ dynes/cm$^2$. For two epoxy joints with width 0.5 mm at radius 1cm and each with a height 2.5 mm, the torsion constant $k_0 = 1.90 \times 10^{12}$ dyne-cm. In this case $k_0 \gg k_\mu$, and a complete treatment of the oscillator mechanics should take potential changes in both $\mu$ and $I_{He}$ into considerations.

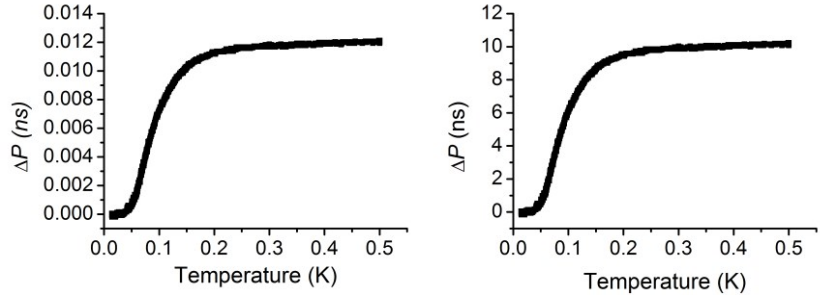

a. Low Mode ($f_{Low}$ = 648 Hz)     b. High Mode ($f_{High}$ = 72.46 kHz)
Fig. 6. Period Shifts of a Compound Torsional Oscillator with Epoxy Joints

## Interpreting Torsional Oscillator Measurements

In Fig. 6 we have plotted the predicted period shifts, based on changes in the shear modulus of the helium sample alone, at the low and high modes of an oscillator with an annulus of radius $r_0 = 1$ cm, width $\Delta r_0 = 0.5$ mm and height $z_0 = 2$ cm, containing the helium sample and an epoxy joint of the same radius and width but height 5 mm. The calculated value for $I_1$ assuming aluminum as the material is 10.6 dyne-cm$^2$. We also assume a value of 80 dyne-cm$^2$ for $I_2$ and $1.5 \times 10^9$ dyne-cm for $k_2$. Using these dimesnsions, the period shift upon filling the cell with solid helium is 535 ns for the low mode. Hence the maximum apparent NCRI fraction due to shear stiffening at this mode is $2.24 \times 10^{-5}$, where we have used the Beamish data set with its 7% decrease in $\mu$. If we had used a $\mu$ variation of 58%, based on ref. 6, the period shifts would be more than 7 times larger and the apparent NCRIF would be $1.86 \times 10^{-4}$.

Next, we consider the case of an internal torsion rod (Fig. 5b), $k_0 \ll k_\mu$ and the mass-loading effect of solid helium on the normal mode frequencies can be safely neglected. The interesting feature of this design is that whereas all oscillators discussed in the previous sections are more sensitive to NCRIs, the floating-core oscillator is much more sensitive to shear-stiffening. The equations of motion can be rewritten as:

$$\begin{pmatrix} k_2 + k_{eff} - \omega^2 I_2 & -k_{eff} \\ -k_{eff} & k_{eff} - \omega^2 I_1 \end{pmatrix} \begin{pmatrix} \theta_2 \\ \theta_1 \end{pmatrix} = 0 \text{, where } k_{eff} = k_\mu + k_0. \quad (39)$$

The floating-core oscillator becomes a simple double oscillator with the coupling between the two oscillators provided by a temperature-dependent $k_{eff}$. The experimentally determined frequencies provide us with an accurate measure of the variation in $\mu$ at low temperatures, since we have the relation:

$$k_{eff} = \frac{2 I_1 I_2 \left( \omega_{High}^2 + \omega_{Low}^2 \right) - k_2 I_1}{I_1 + I_2}. \quad (40)$$

Here $\omega_{High}$ and $\omega_{Low}$ denote the high and low frequency modes with the presence of helium. Expressing (40) in $\mu$, we have shear modulus of solid $^4$He as a function of the normal mode frequencies:

$$\mu = \frac{\Delta r_0}{2\pi r_0^3 z_0} \left[ \frac{2 I_1 I_2 \left( \omega_{High}^2 + \omega_{Low}^2 \right) - k_2 I_1}{I_1 + I_2} - k_0 \right]. \quad (41)$$

In Fig. 7 we have plotted the predicted period shifts for the "floating-core" configuration. The elastic epoxy joints are replaced with a well-defined torsion rod with spring constant $k_0 = 1 \times 10^9$ dyne-cm.

**John D. Reppy, Xiao Mi, Alexander Justin, and Erich J. Mueller**

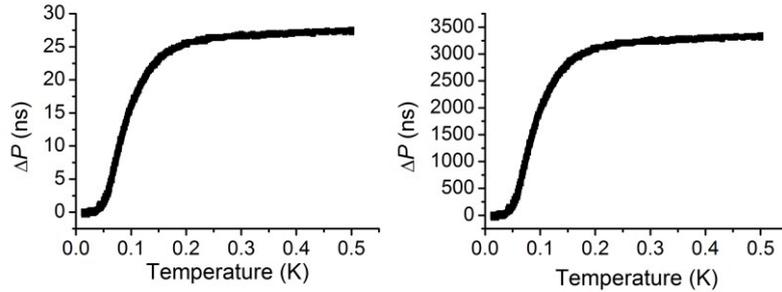

a. Low Mode ($f_{Low}$ = 647 Hz)     b. High Mode ($f_{High}$ = 10.5 kHz)

Fig. 7. Period Shifts of a "Floating-Core" Torsional Oscillator with $k_0 = 1 \times 10^9$ dyne-cm.

The period shifts in such a torsional oscillator are evidently much more pronounced. The period shift upon filling this cell is expected to be 1.81 μs, which gives a maximum apparent NCRIF of 0.0149, or 1.49 percent, assuming a $\mu$ shift of 7%, while we would have an apparent NCRIF of 12.35% for a 58% shift in $\mu$.

**4c. Triple Compound Oscillator**

We will now consider the most complex case, the triple oscillator, which is formed by mounting a compound oscillator on top of a double oscillator. A schematic for this design used in our previous work[14] is shown in Fig. 8.

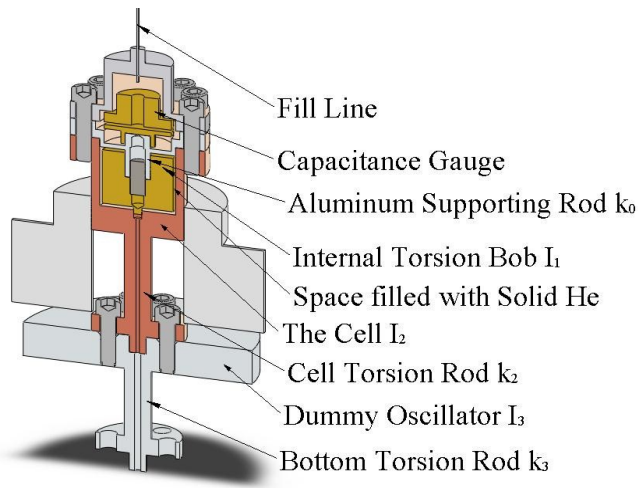

Fig. 8. (Color Online) Triple Compound Oscillator

## Interpreting Torsional Oscillator Measurements

One way to treat the impact of the helium sample is to use the same approach as we have used in the compound oscillator case, writing the equations of motion as:

$$\begin{pmatrix} \left(I_1 + \frac{1}{2}I_{He}\right)\omega^2 - k_1 & k_1 & 0 \\ k_1 & \left(I_2 + \frac{1}{2}I_{He}\right)\omega^2 - k_1 - k_2 & k_2 \\ 0 & k_2 & I_3\omega^2 - k_2 - k_3 \end{pmatrix} \begin{pmatrix} \theta_1 \\ \theta_2 \\ \theta_3 \end{pmatrix} = \begin{pmatrix} 0 \\ 0 \\ 0 \end{pmatrix}.$$

(42)

Here $k_3 = k_0 + k_\mu$, where $k_0$ is the torsion constant of the internal torsion rod and $k_\mu = 2\pi r_0^3 z_0 \mu / \Delta r_0$ is the torsion constant contributed by solid helium. The three normal mode frequencies, in order of increasing magnitude, are denoted by $\omega_-$, $\omega_+$ and $\omega_1$. However, expressing the shear modulus $\mu$ in terms of the normal mode frequencies is very complicated. A simpler approach is justified in the case where the moments of inertia of the helium sample and the internal torsion bob are small compared to the moments of inertia of the cell and the dummy oscillator. We can then treat the system as a double oscillator where a periodic back action torque with amplitude $\tau = \chi(\omega)\theta_2$ acts on the cell in addition to the torque from its own torsion rod. This torque is just that which is required for the angular acceleration of the helium sample and the internal torsion bob during the oscillation of the cell. Then, in terms of a double oscillator we have for the equations of motion:

$$\begin{pmatrix} I_2\omega^2 - k_2 + \chi(\omega) & k_2 \\ k_2 & I_3\omega^2 - k_2 - k_3 \end{pmatrix} \begin{pmatrix} \theta_2 \\ \theta_3 \end{pmatrix} = \begin{pmatrix} 0 \\ 0 \end{pmatrix}.$$

(43)

To estimate the back action torque $\chi(\omega)\theta_2$, we first note that for the two lower modes $\omega_{-/+}$, the motion of the internal torsion bob is in phase with that of the cell and $\theta_1 = \theta_2 + \Delta\theta$. The torque stems from the difference between $\theta_1$ and $\theta_2$, which can be expressed as $\chi(\omega)\theta_2 = k_1\Delta\theta = (k_0 + k_\mu)\Delta\theta$. It also determines the angular acceleration of the internal torsion bob and the sample, which means $\chi(\omega)\theta_2 = -\omega^2(I_{He} + I_1)(\theta_2 + \theta) \approx -\omega^2(I_{He} + I_1)\theta_2$ since $\Delta\theta \ll \theta_2$ which is justified by the fact that $\omega_1 \gg \omega_{-/+}$. Solving for $\chi(\omega)$, we get:

$$\chi(\omega) = -\omega^2\left(I_{He} + I_1\right)\left(1 + \frac{\omega^2\left(I_{He} + I_1\right)}{k_0 + k_\mu}\right)$$

$$= -\omega^2\left(I_{He} + I_1\right)\left(1 + \frac{\omega^2\left(I_{He} + I_1\right)}{k_0 + 2\pi r_0^3 z_0 \mu / \Delta r_0}\right).$$

(44)

**John D. Reppy, Xiao Mi, Alexander Justin, and Erich J. Mueller**

The sensitivity of the periods to variation in mass-loading due to changes in the moment of inertia of the $^4$He solid is:

$$\frac{dP}{dI_{He}} = \frac{dP}{d\chi(\omega)}\frac{d\chi(\omega)}{dI_{He}} \approx -\omega^2 \frac{dP}{d\chi(\omega)} \approx \frac{\delta P}{\delta I_{He}}. \quad (45)$$

We neglected the term $\frac{\omega^2(I_{He}+I_1)}{k_0+2\pi r_0^3 z_0 \mu/\Delta r_0}$ in this process because it is usually much smaller than 1. $\delta P / \delta I_{He}$ is a quantity that can either be calculated from the measured dimensions of the oscillators or determined experimentally, through attaching a mass of known moment of inertia $\delta I$ to the cell and measuring the change in period $\delta P$, essentially simulating a supersolid fraction in the process. For the sensitivity to changes in the shear modulus we have, instead:

$$\frac{dP}{d\mu} = \frac{dP}{d\chi(\omega)}\frac{d\chi(\omega)}{d\mu} = \omega^4 \frac{dP}{d\chi(\omega)}\frac{2\pi r_0^3 z_0 (I_{He}+I_1)^2/\Delta r_0}{(k_0+k_\mu)^2}$$

$$= -\omega^2 \frac{2\pi r_0^3 z_0 (I_{He}+I_1)^2/\Delta r_0}{(k_0+k_\mu)^2}\frac{\delta P}{\delta I_{He}} \quad (46)$$

Here the ratio of sensitivities for small changes in the shear modulus is

$$\left(\frac{\delta P_+}{\delta P_-}\right)_{Shear} = \frac{(dP_+/d\mu)}{(dP_-/d\mu)} = \left(\frac{P_-}{P_+}\right)^2\left[\left(\frac{\delta P_+}{\delta P_-}\right)_{NCRI}\right]. \quad (47)$$

The experimentally determined $\delta P / \delta I_{He}$ values for the -/+ modes and equation (47) enable us to analyze the observed period shifts for the - and + modes based on the expectations for the shear-stiffening scenario or the supersolid scenario. As an example, we consider the data reported in ref.14 for measurements with a triple oscillator. In Fig. 9 we plot the period shift data for the + and – modes referenced to the period values at the lowest temperature near 20 mK. The experimentally determined sensitivities to mass loading for the two modes are $m_+ = \delta P_+/\Delta I = 2.9$ μs/(gm-cm$^2$) and $m_- = 12.4$ μs/(gm-cm$^2$). The ratio of the sensitivities of the two modes to mass loading is $m_-/m_+ = 4.76$. In the figure we also plot an adjusted value, $\Delta P_{-adj} = \Delta P_- /4.76$, for the $\Delta P_-$ data set by dividing by the factor 4.74. If the period shifts observed in this measurement were due to the NCRI phenomenon, then we would expect the adjusted value for the lower mode to agree with the data for the higher frequency mode, since the mass loading effects associated with the NCRI are not expected to depend on frequency. Clearly $\Delta P_{-adj} \neq \Delta P_+$ and we must seek an alternate explanation for the observed period data. If the period shifts arise from the anomaly in the shear modulus,

## Interpreting Torsional Oscillator Measurements

we require an additional adjustment of the low mode period shift data by a factor of $(P_-/P_+)^2 = 1.56$ to account for the effect of the change in the shear modulus. Multiplying $\Delta P_{-adj}$ by this factor bring the lower frequency data set into much better agreement with the higher frequency period shift data. The small remaining disagreement may be due to the presence of a small NCRI term as well as to possible uncertainties in the mass loading calibration.

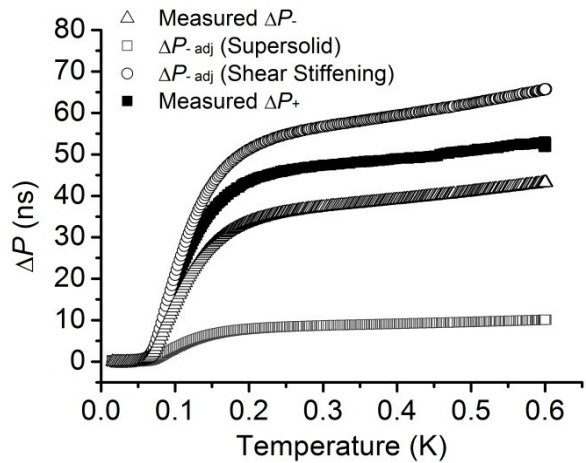

Fig. 9. Triple Oscillator Data

It is very evident that in this particular compound oscillator, most, if not all of the apparent supersolid signal is actually caused by changes in shear modulus of the helium sample.

## 5. CONCLUSION

We have considered the influence of variation of the shear modulus with temperature on the frequencies for a number of torsional oscillators of different designs. There are two basic classes of oscillators: first those with a highly rigid structure and second those in which some significant degree of relative motion between sections of the torsional oscillator exists. In principle, there will always be some contribution to the temperature dependence of a torsional oscillator containing a solid $^4$He sample due to the shear modulus anomaly. The magnitude of the contribution depends on the mode frequency $\omega$ and is proportional to $\omega^2$. Our calculations, based on simple models for the torsional oscillators used for supersolid measurements, show that the changes in the effective moment of inertia of the solid sample due to the anomaly are small for samples confined in rigid cells with a narrow spacing between the walls of the sample volume. In cells with open

**John D. Reppy, Xiao Mi, Alexander Justin, and Erich J. Mueller**

volumes, such as the open cylinder geometry, the apparent NCRI following from the temperature variation of the shear modulus may be detectable. In contrast, for cells where there exists significant relative motion of the constituents, the shear modulus anomaly may add a sizable contribution the observed temperature dependence of the mode frequency and thus lead to an apparent NCRI. Fortunately, multiple frequency torsional oscillators can provide a means for the disentanglement of the effects of the shear modulus anomaly from a true supersolid signal.

**Acknowledgements** The work reported here has been supported by the National Science Foundation through grants DMR-060586, PHY-0758104, and CCMR grant DMR-0520404. The authors would like to thank Michael Vidal for assistance in the construction of cryogenic apparatus.